\documentclass[aps,prl,twocolumn,showpacs,letterpaper,superscriptaddress]{revtex4}
\usepackage{graphicx}
\usepackage{amsmath}

\def\beqn{\begin{eqnarray}}
\def\eeqn{\end{eqnarray}}
\def\beqns{\begin{eqnarray*}}
\def\eeqns{\end{eqnarray*}}
\def\beq{\begin{equation}}
\def\eeq{\end{equation}}
\def\bea{\begin{array}}
\def\ea{\end{array}}

\def\<{\langle}
\def\>{\rangle}

\begin{document}
\title{Quenching of quantum Hall effect and  the role of undoped
planes in multilayered epitaxial graphene}

\author{Pierre \surname{Darancet}}
\affiliation{Institut N\' eel, CNRS/UJF, 25 rue des Martyrs BP166,
38042 Grenoble Cedex 9, France.}
\author{Nicolas \surname{Wipf}}
\affiliation{Institut N\' eel, CNRS/UJF, 25 rue des Martyrs BP166,
38042 Grenoble Cedex 9, France.}
\author{Claire \surname{Berger}}
\affiliation{Georgia Institute of Technology, Atlanta, Georgia, USA.}
\affiliation{Institut N\' eel, CNRS/UJF, 25 rue des Martyrs BP166,
38042 Grenoble Cedex 9, France.}
\author{Walt \surname{de Heer}}
\affiliation{Georgia Institute of Technology, Atlanta, Georgia, USA.}
\author{Didier \surname{Mayou}}
\affiliation{Institut N\' eel, CNRS/UJF, 25 rue des Martyrs BP166,
38042 Grenoble Cedex 9, France.}
\date{\today}

\begin{abstract}

We propose a mechanism for the quenching of the Shubnikov de Haas
oscillations and the quantum Hall effect observed  in epitaxial
graphene. Experimental data show that the scattering time of the 
conduction electron is magnetic field dependent and of the order of
the cyclotron orbit period, \textit{i.e.} can be much smaller than
the zero field scattering time.
Our scenario involves the extraordinary
graphene $n=0$ Landau level of the  uncharged layers that  produces
a high density of states at the Fermi  level. We find that
the coupling between this $n=0$ Landau level  and the
conducting states   of the doped plane
leads to a scattering mechanism having the right
magnitude to explain the experimental data.

\end{abstract}

\maketitle

Graphene is a two dimensional carbon material which takes the form of
a planar honeycomb lattice of $sp^{2}$ bonded atoms. Its two
dimensional character and the linear dispersion relation implies
that, close to the charge neutrality point, the electrons obey an
effective massless Dirac equation. The properties of  electrons in
graphene, deriving from the Dirac equation, are fundamentally
different from those of electrons in standard semi-conductors which
obey  the  Schr\"odinger equation.  A remarkable  example  is the
quantum Hall effect which is quantized with integer plus half values
\cite{novoselov,zhang} and has even been observed  at room
temperature \cite{novoselovQHERT}. Another major fact is the large
electronic coherence \cite{berger,berger2}, that is observed even at
room temperature, and gives the hope to produce devices with new
properties.

Graphene can be produced by mechanical exfoliation of graphite
\cite{novoselov,zhang} which produces either single or few layer
graphene. Another way of producing graphene is by thermal
decomposition of the surface of SiC \cite{berger3,berger,berger2}.  Epitaxial
graphene is attractive because it grows on an insulating substrate,
with a high structural quality and very smooth graphene layers
\cite{Hass}. Both methods produce graphene samples with spectacular
electron coherence properties. Yet magnetotransport remains puzzling
in epitaxial graphene. On one hand, at low magnetic field,  the
experimental  results indicate  electronic mean free paths
  of more than half a micron at 4K \cite{berger2,Wu}. On
the other hand  Shubnikov de Haas oscillations of the
magnetoresistance are weak and the quantum Hall effect is not  observed.
Here we show that  the small amplitude of the Shubnikov de Haas
oscillations is not a signature of a low quality graphene.
We show, by analysing the experimental data, that the results can be explained
by a scattering time of the  conduction electron which is magnetic
field dependent,  and reduced to the order of the
cyclotron orbit period. We argue that
the conducting states in a doped layer can
couple to the zeroth  Landau levels in an undoped layer, which is
on top of the doped one. As a consequence the conducting  electrons
of the doped layer are subjected to a scattering
mechanism that increases with magnetic
field, because the number of states in the zeroth Landau level
increases with magnetic field. At low magnetic field the scattering
time $\tau$ can be long but at stronger field $\tau$ decreases in
such a way that $\omega\tau \simeq 1$, where $\omega$ is the
cyclotron frequency. This forbids the observation of strong
Shubnikov de Haas oscillations. We show also that in some limits the
magnetoresistance increases  linearly with the magnetic field. This
has been observed recently  in epitaxial graphene multi-layers and
the magnitude of the linear magnetoresistance is in quantitative
agreement with our model.

The  letter is organized as follows: we
present  first the experimental results, and show that they can
be explained by a linear dependence of the scattering rate
with the magnetic field. Based on experimental results we
propose a  simple model of electronic structure
and derive the equations satisfied by the Green's functions,
in the Self-Consistent Born Approximation.  Finally we analyse
the electronic structure of this model and show that its
magnetotransport properties  are in good
agreement with the experimental observations.

$\it{Summary~of~previous~experimental~results}$

Electronic properties of epitaxial graphene have been analyzed in
great detail. They clearly point to the existence of a electron doped plane at
the SiC/graphene interface  \cite{Lanzara,rotenberg}
carrying the main part of the current  \cite{berger2}.
  Since the screening length is of $1$-$2$ interlayer spacing, only $2$-$3$ planes are doped, and the
other planes are quasi-neutral, as shown in
infra-red spectroscopy  \cite{sadowski}
and  thus comparatively poor conductors. The experimental evidence that one
 of the doped planes carries the main part of the current suggests 
that the other doped planes have a low conductivity due to the impurities spatial distribution.
The  doped and  quasi-neutral planes all have the characteristics
of an isolated graphene plane \textit{i.e.}
massless  Dirac electrons
\cite{berger2,sadowski,Lanzara,rotenberg}
This was initially surprising since the epitaxial
graphene samples studied in  \cite{berger2} and   \cite{sadowski} consists
of several stacked
graphene planes. Yet, as proven recently, these planes are
rotationally stacked  \cite{Varchon} and the
relative rotation strongly diminishes
the effective electronic  coupling between the planes as compared
to the case of AB (Bernal) stacking
\cite{Lopes,Varchon,Latil,Latil2,ducastelle}.

$\it{Analysis~and~interpretation~of~experimental~results}$

\begin{figure}
    \includegraphics[clip,width=0.5\textwidth]{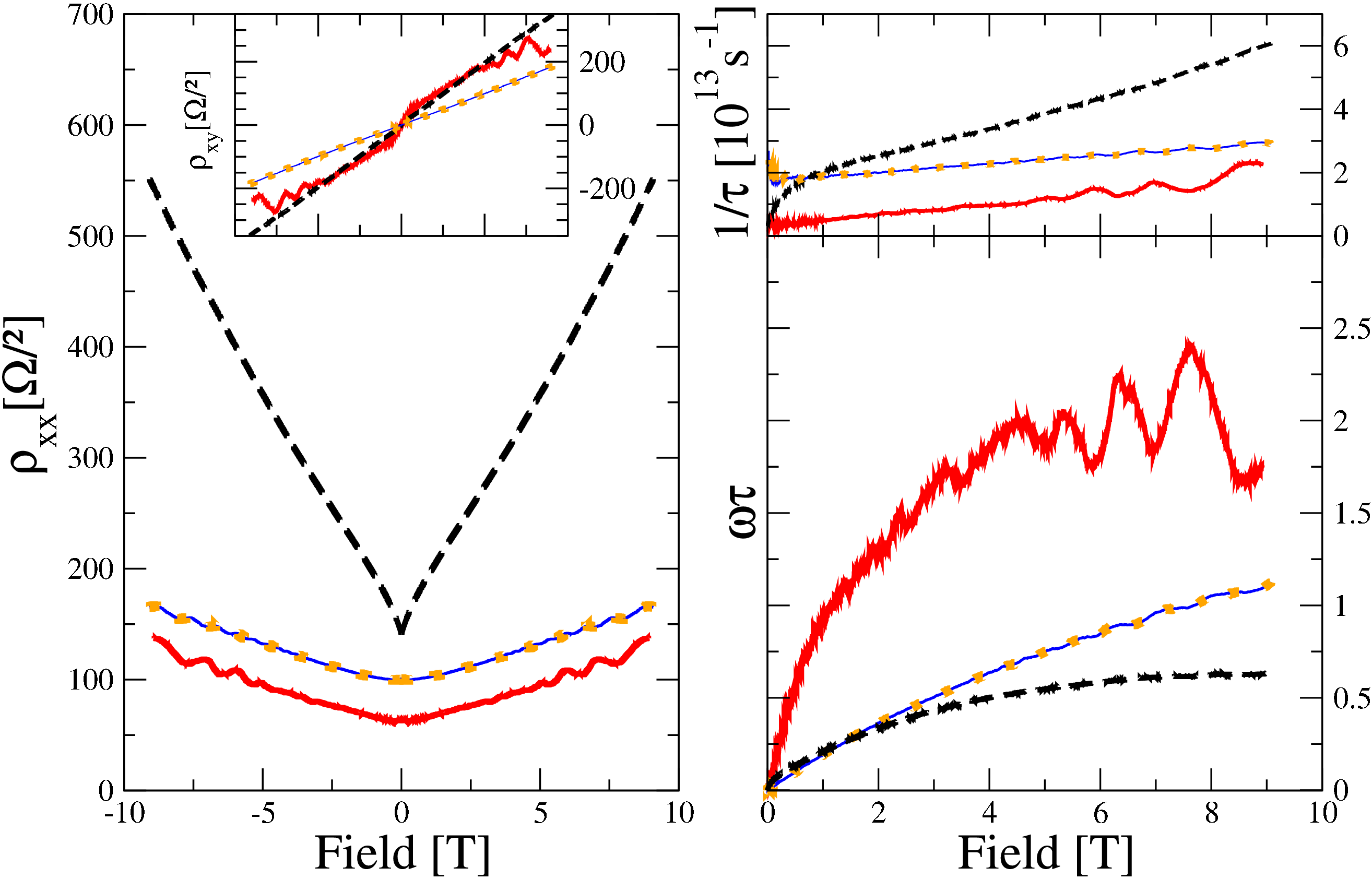}
    \caption{ Left: Experimental values of resistivity as a function of
$B$ for a $100 \mu m \times 1000 \mu m$ (dashed line), a $1\mu m
\times 5 \mu m$ (dashed dotted line), and a $ 0.27 \mu m \times 6 \mu m$
(full line) samples. Main panel: longitudinal  magnetoresistance
$\rho_{xx}$; enclosed panel: transverse magnetoresistance $\rho_{xy}$.
Right: $\rho_{xy}/ \rho_{xx}$ (which is equal to $\omega  \tau$ in
the one band model) (lower panel) and deduced scattering rate (upper
panel). Note that the longitudinal magnetoresistances $\rho_{xx}$
have been already published in the case of the $
0.27 \mu m \times 6 \mu m$ \cite{berger2}
and the $100 \mu m \times 1000 \mu m$ \cite{Wu} samples.}
\label{f3}
\end{figure}

The magnetotransport coefficients $\rho_{xx}$
and $\rho_{xy}$ for three samples of various
widths
($0.27 \mu m$, $5 \mu m$ and $1000 \mu m$) are plotted in figure
(\ref{f3}). Because the quantum corrections to the conductivity are
small \cite{Wu} a semi-classical one band model should be a fair approximation.
The transverse magnetoresistance  $\rho_{xy}$
increases essentially  linearly with the magnetic field $B$ and is
indeed consistent with a semi-classical model.
Yet, the longitudinal magnetoresistance
$\rho_{xx}$ increases linearly with $B$.
From these experimental results, it is possible
to plot the ratio $\omega \tau$ and consequently
to extract the scattering rate $1/\tau$  (figure
(\ref{f3})). This clearly indicates that
the  scattering rate linearly increases with the magnetic field.
We note that there is also experimental  evidence from optical  
measurements that the scattering rate increases with increasing magnetic  
field in these systems as it does in graphite \cite{Basov} . 
Since $\omega$ is given by  $\hbar \omega
=\frac{e B V_{F}}{k_{F}}$, and is so linear with
$B$,
we can deduce from experiments:
\beq
\frac{\hbar}{\tau (B)} \simeq \frac{\hbar}{\tau (B = 0)} + \gamma \hbar \omega
\label{Egenerale}
\eeq
\\
where $\gamma$ is a positive constant. Note that
$1/\tau (B) > 1/\tau(B=0)$, and
$\omega \tau < 1/\gamma$ (with
$\gamma $ equal to $0.4$, $0.45$ and $1.25$
for the present $1\mu m \times 5 \mu m$,  $ 0.27 \mu m \times 6 \mu m$ and
$100 \mu m \times 1000 \mu m$ samples). Moreover, we clearly see
(figure (\ref{f3})) that $\omega \tau$ is directly related to the
system's ability to exhibit Shubnikov de Haas oscillations:
At low $\omega \tau$, the system does not exhibit such oscillations,
while higher values are concomitant with the
appearance of well separated Landau-level in the
electronic structure.
A scattering rate $1/\tau$ linear with $B$ can thus
explain both the unusual magnetoresistance
behaviour and the quenching
of Shubnikov de Haas oscillations.
If one assumes that the coupling between the different planes is negligible
it might seem quite difficult to explain a scattering rate that varies linearly
with the magnetic field. Indeed, the conventional
theory only predicts a quadratic
correction to the magnetoresistance \cite{Abrikosov}.
Yet, a scattering mechanism due
to coupling between the doped and undoped  planes will depend on
the magnetic field since the electronic structure of the undoped
planes depends on the magnetic field.
Even in weak magnetic fields  the Fermi level in
the
nearly neutral planes will be in the low indexes
Landau levels  and  it will be
in the zeroth Landau level as soon as $B >B_{C}= h n_{d}/2e$
where $n_{d}$ is the electron density. Infrared
Landau spectroscopy for the nearly undoped
planes indicate  $n_{d}$ is less than $
10^{10}cm^{-2}$, \textit{i.e.} $B_{C}<<1T$
\cite{sadowski}.
We will see that this scattering mechanism
implies an increasing scattering rate with an
order
of magnitude of $ \Delta ( \hbar/\tau(B) ) \simeq
\hbar \omega $, with
$\gamma$ close to $1$, compatible with the experiments.

$\it{Model~of~electronic~structure}$

We consider  that the perfect rotationally stacked planes are essentially
decoupled since the  coupling between
different states close to the Dirac point is of
the order of  $1 meV$ (compared to
$0.2-0.3 eV$ for A-B stacking) \cite {Varchon}. It is important to
note that the hopping matrix elements between neighboring orbitals of
the two rotated planes are of the order of
$0.2-0.3 eV$, and that the effective
electronic decoupling is due to an averaging specific to the perfect
structure.  This averaging is destroyed by disorder, and consequently
the  defects in one plane  will introduce a scattering in that plane
and a coupling with electronic states in the other plane. Defects
that spread on both planes also introduce in-plane scattering and
interplane coupling. Here  we consider a model with one doped plane
and one   plane which is essentially neutral, for which the zeroth
Landau level is half-filled at all values of the magnetic field. The
two planes are decoupled in the absence of disorder but are coupled
by disorder. In the following plane"1" is the doped plane and
plane "2" the undoped one.

We consider also that the electronic states in plane "1" are coupled
essentially to the zeroth order Landau level of plane "2" only.
Indeed we note that for $B=1T$ the index of the Landau levels of the
doped plane is $n\simeq 30-40$ and there are about $2N_{L}\simeq 15$
Landau levels of the doped plane which are closer to the zeroth
Landau level than to the other Landau levels of plane "2". In the
following we shall treat $N_{L}$ as a large number.

The model for
the Green's functions of the uncoupled and perfect planes "1" and "2"
are $G_{1,0} (z)$ and $G_{2,0} (z)$ :

\beq
G_{1,0} (z)=\sum_{n=-N_{L}}^{n=N_{L}}\frac{N(B) }{z- n\hbar\omega
}~~~,~~~ G_{2,0}(z)=\frac{R N(B)}{z-E_{L0}}
\label{E6}
\eeq

with $\hbar \omega =\frac{e B V_{F}}{k_{F}}$ and $ \frac{N(B)}{\hbar
\omega}=n_{0}=\frac{2k_{F}}{\pi \hbar V_{F}}$. $k_{F}$ is the Fermi
wavevector of the doped plane and $n_{0}$ its density of states at
the Femi energy, without magnetic field. In this model we assume that
the Landau levels of plane "1" are equally spaced by $\hbar \omega $
where $\omega$ is the cyclotron frequency (which  is valid as long as
their index is high) . $E_{L0}$ is the energy of the zeroth Landau
level ($E_{L0} < < N_{L} \hbar \omega$ varies with $B$ so that the
zeroth Landau level stays half-filled), and $R$  is the ratio
between  the number of states in the zeroth Landau level of plane "2"
and in a Landau level of plane "1".  $R$ is equal to $1$ if the two
planes are equivalent (apart from the doping). We can also simulate
the case where the doped plane is coupled to two undoped planes by
taking $R=2$. The degeneracy of a zeroth Landau level of a bilayer in
AB stacking is also $R=2$ \cite{McCannFalko}. Thus different
configurations can lead to different values of $R$ and we let this
parameter free in the following.

In order to treat the effect of in-plane scattering by disorder and
intra-plane coupling by disorder we use a standard Self-Consistent
Born Approximation (SCBA). We introduce the Green's function
$G_{P}(z)$ and the self-energies $\Sigma_{P}(z)$ of  plane "$P$"
($P=1$ or $P=2$). The density  of states per unit surface in plane
$P$  is given by $n_{P}(E)=-1/\pi Im (G_{P}(z = E + i\epsilon))$, where
$\epsilon$ is an infinitely small positive real number. One get:

\beq
G_{P}(z)= G_{P,0} (z- \Sigma_{P}(z))
\label{E5}
\eeq

\beq
\Sigma_{P}(z)= |V|^{2} G_{P'}(z) -i \frac{\hbar}{2\tau_{P,P}}
\label{E2}
\eeq

$P'$ is the plane coupled to plane $P$ ($P'=1$ if $P=2$ and
vice-versa). $V^{2}$ is an average value of the square coupling
between states  in plane "1"  and in plane "2". The terms
$\hbar/2\tau_{P,P}$  represent the effect of in-plane scattering for
each plane $P$  ($\tau_{P,P}$ is the in-plane scattering time ).

   In  (\ref{E6}, \ref{E5}, \ref{E2}) the  two important parameters for
the effect of the interplane coupling  are $R$ and $V$. We analyse
below the regimes of large and small $R$ or $V$. We show that in the
large $V$ limit the experimental magnetic field dependence of the
scattering rate (see figure \ref{f3}) are explained with a
reasonnable value of the parameter R . We emphasize that the physics
described below presents some analogies with the coupling between
localized $d$-orbitals with extended $sp$-states \cite{package}.

   $\it{Large~and~small~R~regimes}$

In this part, we focus on the case $z= E_{L0}+ i\epsilon$. Since the zeroth
Landau level is half-filled, $E_{L0}$ is always close to the Fermi
level. Let us assume that $G_{1}(z=E_{L0} + i\epsilon) \simeq -i\pi n_{0}$ and
thus $n_{1}(E_{L0}) \simeq n_{0}$ (we show below that this
corresponds to the large $R$ limit). After  (\ref{E6}, \ref{E5},
\ref{E2}), this occurs for $|Im\Sigma_{1}(E_{L0}+
i\epsilon)|/\hbar\omega >> 1$.
In this limit, the real parts  $ReG_{2}(E_{L0}+
i\epsilon)$, $ReG_{1}(E_{L0}+ i\epsilon)$,
$Re\Sigma_{2}(E_{L0}+ i\epsilon)$,
$Re\Sigma_{1}(E_{L0}+ i\epsilon)$ are all
negligible.
Using the correspondence $\hbar/\tau=2Im\Sigma$, where $\tau$ is an
electron lifetime, one may write the SCBA equations in a form similar
to the Fermi Golden Rule \textit{i.e.} $ \frac{\hbar}{\tau_{P}}=
\frac{\hbar}{\tau_{P,P}} + 2\pi V^{2} n_{P'}$ with $n_{1} \simeq
n_{0}$ and $n_{2} = RN(B) \frac{2\tau_{2}}{\pi \hbar}$ the densities
of states at $ z = E_{L0}+ i\epsilon$ and then:

\beq
\frac{\hbar}{\tau_{1}} \simeq \frac{\hbar}{\tau_{1,1}} +
\frac{2R}{\pi} \frac{\hbar \omega }{1+\alpha}
\label{E11}
\eeq
\\

with $\alpha = \frac{\hbar / \tau_{2,2}}{2\pi V^{2} n_{0}}$.
Here $\hbar/\tau_{2,2}$ and $ 2 \pi V^{2} n_{0}$ are respectively the
width of the zeroth Landau level due to disorder in plane $2$ and to
coupling with plane "1" in the limit where its density is $n_{0}$.
One sees that if disorder in plane "2" (term $\hbar/\tau_{2,2}$)
increases then the scattering rate $\hbar/\tau_{1}$ decreases.
Indeed  the scattering by  plane "2" is favored by a  strong density
of states in plane "2", whereas the term $\hbar/\tau_{2,2}$ tends to
decrease this density. If  $\alpha >> 1$, the coupling between planes
"1" and "2" has essentially no effect (\textit{i.e.} $\omega \tau_{1}\simeq
\omega \tau_{1,1}$)  but in the opposite limit $\alpha < < 1$ the
scattering rate for electrons in plane "1"  increases linearly with
the magnetic field and $\omega \tau_{1}\leq  \pi /2R$.

We
show now that one  get the regime $G_{1}(z=E_{L0}+ i\epsilon) \simeq -i\pi
n_{0}$ at sufficiently large values of $R$. For simplicity we
consider the limit $\hbar/\tau_{2,2}=\hbar/\tau_{1,1}=0$, but let us
just note that $\hbar/\tau_{1,1}$ and
$\hbar/\tau_{2,2}$ have opposite effects,
since $\hbar/\tau_{1,1}$ favors the large $R$ regime  while
$\hbar/\tau_{2,2}$ favors a small $R$ regime. The SCBA equations can
be written with dimensionless quantities
${F}_{P}= \tilde{F}_{P}(\frac{z}{\hbar\omega},
\frac{E_{L0}}{\hbar\omega},\frac{2 \pi V^{2}n_{0}}{\hbar\omega},R)$,
where ${F}_{P}= \frac{G_{P}(z)}{n_{0}}$ or
${F}_{P}=\frac{\Sigma_{P}(z)}{\hbar\omega}$
At  $z=E_{L0} + i\epsilon$ one get for  $G_{1}(z)$, after (\ref{E6}, \ref{E5},
\ref{E2}) $G_{1}(z)= G_{1,0} (z- R \frac{N(B)}{G_{1}(z)})$. This
equation  is independent of the coupling parameter $V$, and in that
case $n_{1}/n_{0}$ and $2 Im\Sigma_{1}/\hbar\omega=1/\omega\tau_{1}$
are functions only of $E_{L0}/\hbar\omega$ and $R$.

\begin{figure}
     \includegraphics[clip,width=0.5\textwidth]{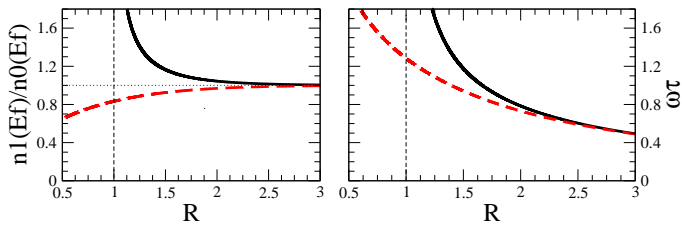}
    \caption{Left: value of $n_{1}(E_{F})/n_{0}$ as a function of $R$
for $E_{L0}/\hbar\omega = 0$ (full line) and  for
$E_{L0}/\hbar\omega = 1/2$ (dashed line). Right: Same for the ratio
$\omega \tau$.
    }
    \label{f1}
\end{figure}

We consider  $E_{L0}/\hbar\omega = 0$ and $E_{L0}/\hbar\omega =1/2$
for which by symmetry $E_{L0}=E_{F}$ and the real parts of the
self-energies and Green's functions are zero. The result is shown in
figure (\ref{f1}). For large $R$ typically $R \gtrsim 1.5-2$ one has
$n_{1}(E_{F})/n_{0}\simeq 1$ which is the criterion for the strong
scattering regime. In that case one recovers the strong scattering
limit $\hbar/\tau_{1}\simeq 2R/\pi \hbar\omega$
that is $\omega\tau_{1}=\pi/2R$.
Note also that in the large $R$ regime  the results are essentially
independent of $E_{L0}/\hbar\omega$. At $R=1$ (\textit{i.e.} not in the large
$R$ regime)  and for  $E_{L0}/\hbar\omega=0$  the density of states
$n_{1}(E_{F})$ diverges. Indeed as soon as $R<1$ there are less
states in the zeroth Landau level of plane "2"  than in Landau levels
of plane "1".  This means that there exist uncoupled states of the
Landau level of plane "1" and thus an infinite density $n_{1}(E_{F})$
as soon as $R < 1$.

\begin{figure}
    \includegraphics[clip,width=0.5\textwidth]{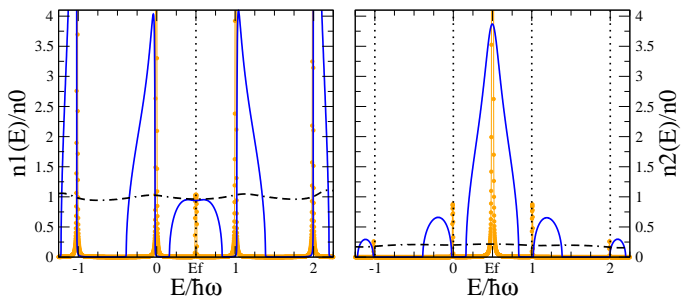}
    \caption{ Left: Dimensionless density of states in plane "1"
($n_{1}(E)/n_{0}$) as a function of energy for different coupling
$V^{2} n_{0} / \hbar\omega =$ $1$ (dashed dotted line); $6.10^{-2}$
(full line); $1.10^{-3}$ (dotted line) and $R=2$, $E_{L0}/\hbar\omega
= 1/2$ . Right: same for the dimensionless density of states
($n_{2}(E)/n_{0}$) in plane "2".}
\label{f2}
\end{figure}

$\it{Large~and~small~V~regimes}$

We define the large and small $V$ regimes respectively  by  $ 2 \pi
V^{2} n_{0}/\hbar\omega >> 1$ and $ 2 \pi V^{2} n_{0}/\hbar\omega <<
1$. After the dimensional analysis  the width $W$  of the zeroth
Landau level of plane "2" and thus the energy range on which the
electronic structure is modified by the coupling satisfies
$W/\hbar\omega = \tilde{W}
(E_{L0}/\hbar\omega,V^{2}n_{0}/\hbar\omega,R)$.

In the large $V$ regime, as long as $|z-E_{L0}|/\hbar\omega \leq 2
\pi V^{2}n_{0}/\hbar\omega$  the term $|z-E_{L0}| << |V^{2}
G_{1}(z)|$ and one recovers $G_{1}(z)= G_{1,0} (z- R
\frac{N(B)}{G_{1}(z)})$. The dimensionless $G_{1}(z)/n_{0}$ and
$\Sigma_{1}(z)/\hbar\omega$ depend on $z/\hbar\omega$ and $R$ but not
on $E_{L0}/\hbar\omega$  and $2 \pi V^{2}n_{0}/\hbar\omega$. Note
that the periodicity $\hbar\omega$ of $G_{1,0}(z=E+i\epsilon)$
(\ref{E6})
implies the same periodicity for $G_{1}(z=E+i\epsilon)$ and
$n_{1}(E)$ in this limit. For the density of states in plane "2" one
has after (\ref{E6}, \ref{E5}, \ref{E2})
$n_{2}(E)/n_{0}=(R/\pi^{2})(\hbar\omega/V^{2}n_{0})(n_{0}/n_{1}(E))$.
Thus $n_{2}(E)/n_{0}$ presents the same periodicity (within the range
$W$) and is very small. The spectral weight $RN(B)$ of the zeroth
Landau level spreads on a width $W$ and thus since
$n_{2}(E)/n_{0}\simeq (R/\pi^{2})(\hbar\omega/V^{2}n_{0})$ one has
$W/\hbar\omega \simeq \pi^{2} V^{2}n_{0}/\hbar\omega$. Finally in the
small$V$ regime $2 \pi V^{2}n_{0}/\hbar\omega <<1$ the width
$W/\hbar\omega <<1$ and the Landau levels of plane "1" and "2" are
only slightly hybridized (see Figure (\ref{f2})).

$\it {Magnetotransport}$

When the density of states is uniform on an
energy scale $W>>\hbar\omega$,
which is the case in the large $R$ and $V$ regime (see  figure (\ref{f2})),
we can apply the semiclassical theory of transport.
The scattering time $\tau_{1}$ is given by  $\hbar/\tau_{1}
=\hbar/\tau_{1,1}+ 2 \hbar \omega
R/(\pi(1+\alpha))$ where $\hbar/\tau_{1,1}$ is
the in-plane
scattering rate. As long as $\alpha$ is not too
large  we expect the term $2 \hbar \omega
R/(\pi(1+\alpha))$ to be of order $\hbar \omega$
and the model is consistent  with the
experimental results presented above
(see figure \ref{f3} and equation
\ref{Egenerale}).  Shubnikov de Haas
oscillations can occur when the field dependent
scattering studied here is
destroyed,  that is   for $\alpha >> 1 $. This can be
due for example to disorder in plane "2"
or even to a confinement effect as in a ribbon of
finite width \cite{guineaconfi}, as in figure (\ref{f3}).
Shubnikov de Haas oscillations can thus be
enhanced by disorder or by confinement effects.
This
spectacular effect is clearly observed in experiments
\cite{berger}.

To conclude we have shown that magnetotransport experiments on epitaxial
graphene are consistent with a scattering time that is  magnetic field
dependent and  is reduced to the order of the cyclotron period.
This explains the unusual variation of the magnetoresistance, 
the quenching of the Shubnikov  de Haas oscillations and of the quantum Hall effect .
The magnitude of the  field dependent scattering time is consistent
with a mechanism where the conducting electrons of
the doped plane  are scattered due to their coupling
with the zeroth Landau level of the undoped planes .

We thank X. Wu,  L. Magaud, V. Olevano, G. Trambly de Laissardi\`ere and F.
Varchon for many stimulating exchanges.  One of us (D.M.)  thanks
also Pascale Lefebvre. We acknowledge a travel grant from CNRS-DREI, and NSF funding  under grant 4106A68 and the W.M. Keck foundation.

\end{document}